# AN INTEGRATED ENTERPRISE ACCELERATOR DATABASE FOR THE SLC CONTROL SYSTEM

T. Lahey, J. Rock, R. Sass, H. Shoaee, K. Underwood
SLAC, Stanford, CA 94309, USA

Abstract

Since its inception in the early 1980's, the SLC Control System has been driven by a highly structured memory-resident real-time database. While efficient, its rigid structure and file-based sources makes it difficult to maintain and extract relevant information. The goal of transforming the sources for this database into a relational form is to enable it to be part of a Control System Enterprise Database that is an integrated central repository for SLC accelerator device and Control System data with links to other associated databases.

We have taken the concepts developed for the NLC Enterprise Database and used them to create and load a relational model of the online SLC Control System database. This database contains data and structure to allow querying and reporting on beamline devices, their associations and parameters. In the future this will be extended to allow generation of EPICS and SLC database files, setup of applications and links to other databases such as accelerator maintenance, archive data, financial and personnel records, cabling information, documentation etc. The database is implemented using Oracle 8i. In the short term it will be updated daily in batch from the online SLC database. In the longer term, it will serve as the primary source for Control System static data, an R&D platform for the NLC, and contribute to SLC Control System operations.

## 1 DESIGN DECISIONS

The decision to stay with Oracle was made easily and early: it is in widespread use by other laboratories; there was substantial in-house expertise; the relational model is well known and understood; experience with OO databases has produced ambiguous results. These factors eliminated the possibility of using another relational or OO database.

Selection of the relational methods used to represent device attributes and their relationships was a more difficult issue [1]. To date, each laboratory has developed their own schema [2] with some, like SNS, building on previous work done at CERN [3]. We had already developed a prototype design for the NLC Enterprise Database [4]. Building on that work, we extended the model to incorporate the device attributes and relationships in the SLC database. In the process, we considered two methods of handling device attributes.

*Attributes as named rows:* In this model, each attribute value and its name are stored as a row in a relational table. This is the method used by the PEPII device database at SLAC [5]. Its main advantage is flexibility – the task of adding devices and/or attributes is a simple matter of adding data, not changing structure. However, since each attribute is a row, queries tend to retrieve large numbers of rows and attribute tables can become large, adversely affecting performance. In addition, this model does not allow optimal use of database features such as constraints and joins, nor the standard set of database query and reporting tools.

*Attributes as table columns:* This is the "standard" way to construct a normalized relational database. Each attribute is a column in a table. The advantages of this approach are clarity of structure, maximal usage of relational features (e.g. constraints and joins) and tools, and faster queries returning fewer rows. Of course, adding devices and attributes requires changes in database structure, a disadvantage in a rapidly changing R&D environment.

We decided on a standard relational model but organized the tables into an OO-like class hierarchy. In addition, every "object" in the database has a unique ID and can be linked to any other object. We can use the standard Oracle tools for design and reporting, and most integrity checking and relationships are implemented in the database structure. The hierarchy of tables allows us to define devices in an Object Oriented-like manner and isolate database changes for new attributes and devices. At the top of this hierarchy is a Linkable Objects table. Using this table we can also link devices together in arbitrary ways. However, addition of new devices and attributes still requires database changes, a maintenance issue that we will need to deal with as development continues.

## 2 DESIGN DETAILS

### 2.1 SLC Database Characteristics

The SLC database consists of a set of flat structures called primaries. Each primary has a set of related data and pointers or references to other primaries. Special programs are required to traverse links and find relationships. Since there is no language or standard way to traverse these links, it can be difficult to extract interesting data from this complex web of connections. All sources are kept in versioned text files. The SLC database presently contains over 1 million data items, each equivalent to an EPICS channel.

### 2.2 Main Relational Tables

Figure 1 shows a few of the main relationships in the database. Every "object" in the database has an entry in the Linkable Objects table. This includes all devices and signals. Everything in the database is a subtype of a Linkable Object. The Accelerator Device table inherits from Linkable Objects and is the parent for most of the usual devices. Controller inherits from Accelerator Device and is the parent for all of the control devices in the system, a large subtype being magnets. One subtype of the Magnets table is Quads. Many controllers also have power supplies associated with them.

IOCs, Crates and Modules inherit directly from Accelerator Device. There are one-to-many mappings of IOCs to Crates to Modules to Signals.

Standalone analog and digital signals inherit directly from the Linkable Objects table. Non-physical information like TWISS parameters, polynomials etc. also inherit directly from the Linkable Objects table.

Database views that join data throughout the hierarchy hide some of the complexity from applications or users, allowing them to obtain data without explicitly joining many tables

In the present design, all of the links e.g. magnet to power supply are explicitly implemented in the tables using database constraints. In the near future, we would also connect Linkable Objects into a hierarchy of related devices, thus a magnet, its power supply (now explicitly connected) temperature thermocouple and BPM would all be associated at the Linkable Object level. Because the SLC database only contains some of this associative information, this aspect of the database has not yet been implemented.

Associations between Linkable Objects would be contained in a single table that could be traversed using an sql select statement with a connect-by clause.

This makes possible a "parts-explosion" hierarchy of devices. The cost of this approach is maintenance of the links and increased query time.

Figure 1  Main Relational Tables

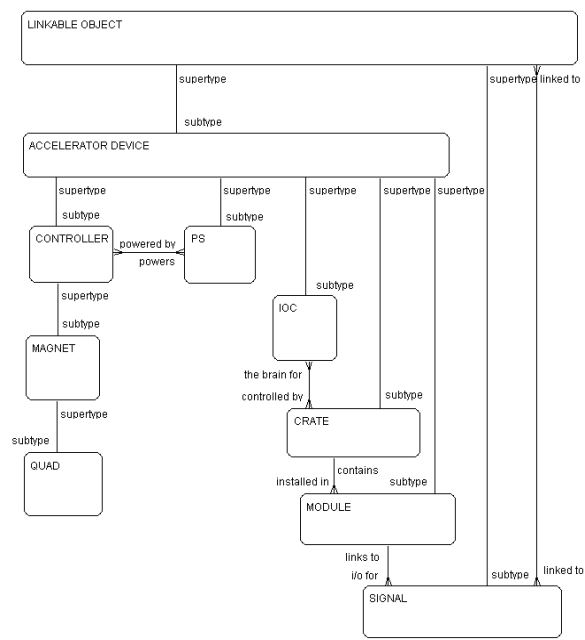

The database presently has about 130 tables. These store most of the information in the SLC database.

## 3 DATABASE LOADING AND MAINTENANCE

The database is maintained on central SLAC UNIX servers. Updating of the database takes place in three stages:

1. Extract the SLC data to ASCII files on VMS.
2. Load into flat Oracle tables in the UNIX database.
3. Populate the full relational structure in the UNIX/Oracle database from the flat tables.

Updating of the relational structure from the SLC database will be initiated automatically whenever there are changes to the SLC database structure. A single script is executed on VMS that starts a series of actions that run on both VMS and UNIX. Because reloading the entire database structure from scratch can take up to two hours, the scripts detect changes since the last time and just update data that has changed. Of course we also have a reload everything from scratch option as well.

## 4 CURRENT TASKS

Our first application of the relational database is a replacement for an application called CAMDMP that extracts and reports CAMAC allocation information in the SLC database.

We are in the process of testing the relational model using queries from SLC applications, and will modify and enhance the structure as needed to accommodate these applications.

The Aida project, [6] which is a CORBA based method to access any data in the Control System, uses an extension of the database to access SLC data.

## 5 FUTURE PLANS

Adding the linking of devices at the level of the Linkable Objects table will provide us with a complete hierarchy of device relationships that is not presently represented in the SLC database.

Other EPICS installations have done various implementations to represent an EPICS database in a relational model. Since EPICS has been loosely integrated into our existing Control System and is used to control several subsystems, we must also integrate it into this database.

Eventually we want to replace the existing text files with this database as the configuration repository for the SLC Control System.

As this database becomes more central to the operation of the SLC Control System, we'll need to develop a set of user interfaces and specialized tools for data retrieval and database maintenance.

## 6 DESIGN SHARING?

Since most labs are using Oracle and we all have a similar problem to solve, can we share basic relational designs between labs? Do the differences overwhelm the similarities? How and where are schemata published? How modified? Who's in charge?